\documentclass[lettersize,journal]{IEEEtran}
\usepackage{amsmath,amsfonts}
\usepackage{algorithmic}
\usepackage{array}
\usepackage[caption=false,font=normalsize,labelfont=sf,textfont=sf]{subfig}
\usepackage{textcomp}
\usepackage{url}
\usepackage{verbatim}
\usepackage{graphicx}
\usepackage{cite}
\usepackage{color}
\usepackage{xcolor}
\usepackage{dblfloatfix}

\usepackage[linesnumbered,ruled, boxed, lined]{algorithm2e}

\usepackage{enumitem}

\begin{document}

\title{ {\color{black} Multi UAV-enabled Distributed Sensing: Cooperation Orchestration and Detection Protocol } }

\author{Xavier A. Flores Cabezas,~\IEEEmembership{Student Member,~IEEE,}
Diana P. Moya Osorio,~\IEEEmembership{Senior Member,~IEEE,}
Markku Juntti,~\IEEEmembership{Fellow Member,~IEEE,}\vspace{-2em}
\thanks{X. A. Flores Cabezas, Diana P. Moya~Osorio, and M. Juntti are with the Centre for Wireless Communications (CWC), University of Oulu, Finland (e-mails: \{xavier.florescabezas;diana.moyaosorio;markku.juntti\}@oulu.fi).}
\thanks{This research was supported by the Research Council of Finland (former Academy of Finland) 6G Flagship Programme (Grant Number: 346208) and project FAITH (Grant Number: 334280).}
}

\maketitle

\begin{abstract}
This paper proposes an unmanned aerial vehicle (UAV)-based distributed sensing framework that uses orthogonal frequency-division multiplexing (OFDM) waveforms to detect the position of a ground target, and UAVs operate in half-duplex mode. A spatial grid approach is proposed, where an specific area in the ground is divided into cells of equal size, then the radar cross-section (RCS) of each cell is jointly estimated by a network of dual-function UAVs. For this purpose, three estimation algorithms are proposed  employing the maximum likelihood criterion, and digital beamforming is used for the local signal acquisition at the receive UAVs. It is also considered that the coordination, fusion of sensing data, and central estimation is performed at a certain UAV acting as a fusion center (FC). Monte Carlo simulations are performed to obtain the absolute estimation error of the proposed framework. The results show an improved accuracy and resolution by the proposed framework, if compared to a single monostatic UAV benchmark, due to the distributed approach among the UAVs. It is also evidenced that a reduced overhead is obtained when compared to a general compressive sensing (CS) approach.
\end{abstract}


\section{Introduction}\label{sec:Introduction}

Sensing services, enabled by the emerging concept of integrated sensing and communications (ISAC) in future perceptive mobile networks (PMN), will not only allow for a more efficient use of network resources, but a further exploration of the utility of cellular systems by supporting several use cases~\cite{9945983,wpaper:Nokia}. Sensing tasks include detection, localization and tracking, imaging, and recognition~\cite{9945983}. Localization can be interpreted as a parameter estimation problem, with the the mean squared error (MSE)~\cite{8252808} or the Cr\'{a}mer-Rao bound (CRB)~\cite{9814670} as its performance metrics. For improving localization accuracy, the coherent joint transmission and reception and centralized signal processing in distributed sensing settings can be exploited, which also alleviates the need of full-duplex operation at sensing nodes~\cite{9729751}.

{\color{black}

Distributed sensing over cellular networks has been conceptualized to give shape to the concept of PMNs, which has been given increasing attention~\cite{art:Heath_PMNMain,art:Song_PMNMag,art:Heath_PMNFramework,art:Zhang_PMNUplink,art:Song_PMNLeak}. Particularly, in~\cite{art:Song_PMNMag}, a comprehensive summary on the opportunities and challenges raised by PMNs is provided. In \cite{art:Heath_PMNFramework}, a framework for distributed sensing over a PMN is addressed by employing orthogonal frequency division multiple access (OFDMA) and multi-user MIMO. This framework includes uplink, downlink active, and downlink passive sensing as types of sensing and compressive sensing (CS) as the method for parameter estimation. On the other hand, synchronization has proved to be a great challenge in the context of PMNs, which has been considered in\cite{art:Zhang_PMNUplink}. Therein, a technique to remove the effect of the timing and phase offsets at the receiver side is proposed when the sensing is performed in the uplink. This is attained by correlating the outputs of receive antennas followed by a high-pass filtering. 
}

It is also evident that leveraging more flexible nodes, such as unmanned aerial vehicles (UAVs), can expand the capabilities of distributed sensing systems by providing more degrees of freedom, as already investigated in~\cite{art:Meng_UAVsISAC} \cite{art:Zhang_MultiUAVSensing}\cite{Chen_JSACUAVSystem}{\color{black}
\cite{art:Zhang_UAVISAC}.
}
Particularly, in~\cite{art:Meng_UAVsISAC}, the beamforming, user association, position and sensing schedule of a UAV, deployed to provide communication with users while sensing targets, are optimized to maximize the sum rate of communications while maintaining sensing requirements. Moreover, in~\cite{art:Zhang_MultiUAVSensing}, the deployment of multiple UAVs is considered for carrying out the detection of multiple targets. Therein, sensing and communication tasks are performed simultaneously by considering different beams, and the target distance estimates are aggregated at a ground base station designated as the fusion center. 
In~\cite{Chen_JSACUAVSystem}, a full-duplex UAV-based ISAC system is proposed, where multiple UAVs perform local sensing while considering reflections from other UAVs as clutter. In this work, the area-based metric named upper-bound average cooperative sensing area (UB-ACSA) is proposed, as the area of sensing coverage in which a given probability of detection and probability of false alarm for sensing are guaranteed, while also guaranteeing a given outage probability for communications. 

A common approach for performing localization of targets, over a certain region, is to discretize a certain domain of interest into a grid made up of several cells. Thus, the localization of targets is made based on the closest cell that corresponds to the estimated target parameters, which is referred to as on-grid methods~\cite{art:Baquero_mmWaveAlg}\cite{art:Zhang_DBR}{\color{black}\cite{art:Malioutov_1Dongrid}\cite{art:Amiri_2Dpostproc}\cite{art:Liu_Grid}\cite{art:Li_OffGridTargLoc}
}. For instance, in~\cite{art:Baquero_mmWaveAlg}, an on-grid approach is employed for target tracking, where sparse recovery methods are utilized over a discretized range and angular grids. Moreover, in~\cite{art:Zhang_DBR}, an on-grid approach is employed for the estimation of delays of targets from echoes of orthogonal frequency-division multiplexing (OFDM) waveforms, where the delay range is discretized, and a one-dimensional (1D) multiple measurement vector (MMV) CS technique is utilized for estimation. 

However, on-grid methods usually entail off-grid errors, since targets are rarely located exactly at the grid points. To deal with such errors, off-grid methods can be used to refine the estimate of the position of the target by considering the deviations of the target parameters from the grid points \cite{art:Abtahi_OffGridMIMO}\cite{art:Park_2DOffgrid}{\color{black}\cite{art:Feng_postproc}.
}For instance, a weighted average off-grid approach is proposed in~\cite{art:Abtahi_OffGridMIMO}, where an on-grid sparse recovery algorithm is applied to jointly obtain the estimates of each cell point and the weighted average coefficients corresponding to each cell point, to perform an off-grid approximation through a weighted average of the positions of the cells. Similarly, an off-grid method is used in~\cite{art:Park_2DOffgrid} to estimate the direction of arrival from multiple radiating sources, where the estimation problem is formulated as a block-sparse CS framework capable of distinguishing closely located sources with high resolution. In~\cite{art:Feng_postproc} an off-grid post-processing technique for multiple-target localization based on received signal strength (RSS) measurements is presented, where a weighted average is performed with the cells that have a value above a certain threshold to obtain an off-grid estimate of the locations of the targets.


\vspace{-1em}
{\color{black}
\subsection{Contributions}
Considering that a 1D delay-based grid introduces ambiguity along ellipses that exhibit the same total delay of the reflections from the scatterers, grids based on delay and angle-of-arrival (AoA) are used to handle the ambiguity in the delay domain by processing on the AoA domain. For the two-dimensional (2D) scenarios, a simple way to perform estimation is to employ 2D grids~\cite{art:Baquero_mmWaveAlg}\cite{art:Baquero_OFDM},\cite{art:Xie_2DSensing}. Moreover, in three-dimensional (3D) scenarios, the grids need to be augmented, for example, to a delay-elevation-azimuth 3D search. This search would directly increase the complexity of the estimation as investigated in \cite{art:Wang_3DSensing},\cite{art:Zhen_CS_3DSensing}. To deal with this dimensional increase, this work performs sensing over a 2D spatial grid in a 3D environment, which gives a simple and straightforward baseline for distributed sensing over different aerial nodes. 

Given the benefits of using UAVs for sensing purposes, the protocol described in this work utilizes a network of UAVs performing distributed sensing to locate a point-target in the ground. The UAVs operate in half-duplex mode, and they are coordinated by another UAV, a fusion center (FC), which acts as a coordinator for the protocol, as well as the central estimator that gathers local statistics of the received signals from the UAVs to estimate the position of the target. In contrast to CS techniques that allow for high resolution estimation of target parameters, herein estimation techniques based on the maximum likelihood estimation (MLE) criterion are considered. The reasoning behind this is that CS techniques present a complexity that can be prohibitive in scenarios of continuous monitoring, and the overhead introduced for a distributed sensing approach will tend to be large. However, the proposed techniques require a small overhead for transmission of the local statistics to the FC compared to a general CS approach. Moreover, considering the promising results provided by off-grid methods, this work explores further resolution enhancements by applying on-grid and off-grid refinement techniques, as well as different fusion mechanisms.
}

{\color{black}

To summarize, the main contributions of this paper are the following.
\begin{itemize}
    \item A novel distributed sensing framework is proposed for target detection using multiple UAVs operating in half-duplex mode, and the detection is based on a spatial grid.
    \item Three distributed target estimation methods are proposed for the central estimation of the position of the target at the FC, based on the MLE criterion over OFDM frames, which reduce the amount of overhead compared to a general CS approach.
    \item To enhance the detection accuracy of our framework, an augmented spatial mixed grid approach and a threshold-based weighted average post-processing approach are proposed.
    \item Exhaustive Monte Carlo simulations are performed to demonstrate the gain on accuracy introduced by the consideration of multiple UAVs over a single UAV benchmark, for different system parameters.
\end{itemize}

\subsection{Paper outline and notations}
The rest of this paper is organized as follows. The system model and signal model are introduced in Section~\ref{sec:SysModel}. The proposed distributed sensing protocol is detailed in Section~\ref{sec:Protocol}. The on-grid and off-grid position estimation of the target is explained in Section~\ref{sec:posEst}. Results are shown for the distributed sensing protocol in Section~\ref{sec:Results}. Finally, conclusions are drawn in Section~\ref{sec:Conclusions}.

\textit{Notations.} In this paper scalar variables are denoted by lowercase, italic letters (e.g. $z$), column vectors are denoted by lowercase bold letters (e.g. $\mathbf{w}$), matrices are denoted by uppercase bold letters (e.g. $\mathbf{H}$) and sets are denoted by uppercase calligraphic letters (e.g. $\mathcal{P}$). Also, $|z|$ represents the modulus of complex scalar $z$, $|\mathcal{P}|$ represents the cardinality of set $\mathcal{P}$, $|| \cdot||_2$ represents the $L^2$ norm of a vector, $|| \cdot||_\infty$ represents the $L^\infty$ norm of a vector, $\odot$ represents the Hadamard product, $\otimes$ represents the Kronecker product, $z^*$ represents the conjugate of complex scalar $z$, $\mathbf{v}^H$ represents the conjugate transpose of complex vector $\mathbf{v}$, and $\mathcal{R}\{\cdot\}$ represents the real part of the complex argument (scalar, vector or matrix).


}

\section{System Model}\label{sec:SysModel}

\begin{figure}[bt]
    \centering
    \includegraphics[width=0.7\linewidth]{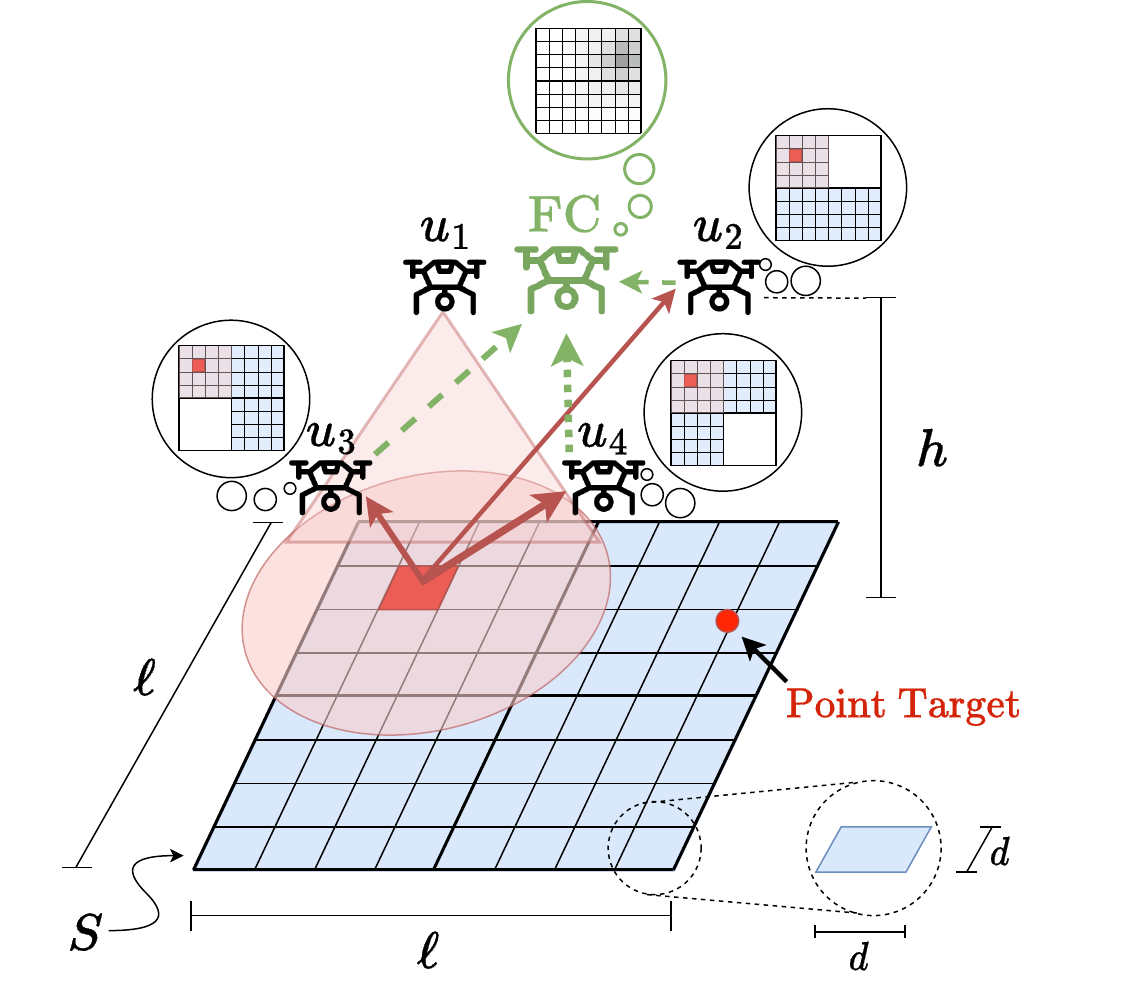}
    \caption{System model.\vspace{-1em}}
    \label{fig:sysModel}
\end{figure}

Consider the system depicted in Fig.~\ref{fig:sysModel}, where a single point-like target of radar cross-section (RCS) $\sigma_{\mathrm{T}}$ is positioned on a square area $S$ of $\ell$ meters of side length. In this system, $U$ UAVs are deployed at a common altitude $h$ and are coordinated to perform distributed sensing in order to locate a ground target over $S$. Each UAV $u \in \mathcal{U}$, is positioned at coordinates $\mathbf{r}_u = [x_u,y_u,h]^T$, with $\mathbf{r}_u\in\mathbb{R}^{3\times 1}$, $|\mathcal{U}|=U$ and $\mathcal{U}$ as the set of all UAVs. It is assumed that the total ground area of interest has an RCS $\sigma_{\mathrm{G}}$, and that it is uniformly spread across it.


Similar to~\cite{Chen_JSACUAVSystem}, it is assumed that each UAV has two arrays of antennas, a square uniform planar array (UPA) for sensing (mounted facing downward)  and a uniform linear array (ULA) for communications (mounted horizontally). It is assumed that radar and communication links use their own dedicated frequencies, so they do not interfere with each-other. The square UPA consists of $N$ isotropic antenna elements spaced $\lambda/2$ from each-other, where $\lambda=f_0/c_0$ is the wavelength of the signal, $f_0$ is the frequency of the signal, and $c_0$ is the speed of light. There is also a fusion center UAV (FC) that performs information fusion and coordination tasks.

\begin{figure}[bt]
    \centering
    \includegraphics[width=0.85\linewidth]{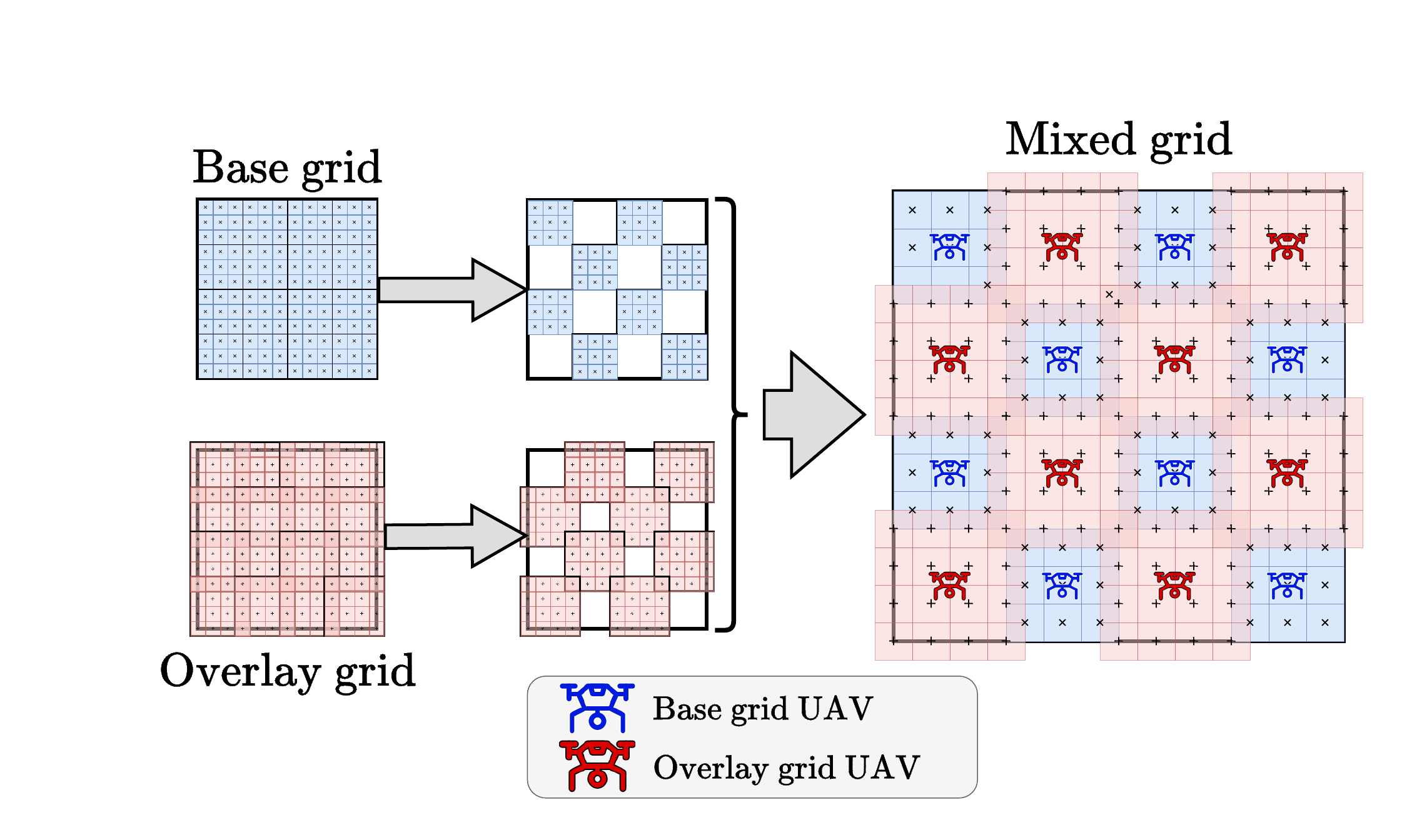}
    \caption{Mixed grid formation.}
    \label{fig:mixedGrid}
    \vspace{-1em}
\end{figure}

For the sensing process, it is considered that the total area $S$ is sectioned into two grids, base grid and overlay grid, and then combined into a mixed grid as shown in Fig.~\ref{fig:mixedGrid}. The base grid is composed of $L\times L$ square cells with dimensions $ d \times d $ such that $d = \ell/L$, while the overlay grid is a shifted version of the base grid, with cells of the same size $d$ shifted horizontally and vertically by $d/2$. Defining the set of all cells as $\mathcal{P}$ with $|\mathcal{P}| = P$ total cells, every cell is characterized by its middle point $p\in\mathcal{P}$ at coordinates $\mathbf{r}_p =[x_p,y_p,0]^T$, and the point $p^*$ represents the target at coordinates $\mathbf{r}_{p^*} =[x_{p^*},y_{p^*},0]^T$. 

OFDM frames consisting of $M_s$ OFDM symbols, each consisting of $N_c$ orthogonal subcarriers are used for the illumination of the cells. Assuming UAV $u\in\mathcal{U}$ illuminates cell $p$ employing a transmit beamformer $\mathbf{w}_{\mathrm{TX}}(p)\in\mathbb{C}^{N\times 1}$, and its reflections are received at UAV $u'\in\mathcal{U}\setminus\{u\}$, which employs the receive beamformer $\mathbf{w}_{\mathrm{RX}}(p)\in\mathbb{C}^{N\times 1}$, the received symbol $\Hat{c}_{k,l}$ corresponding to the OFDM symbol $l\in\{0,...,M_s-1\}$ in subcarrier $k\in\{0,...,N_c-1\}$ considering ground reflections from a discrete number $P$ of reflectors is given as
{\small\begin{align}
\label{eq:symbol_0}    \Hat{c}_{k,l} &= \sum_{p'=1}^{P} \mathbf{w}_{\mathrm{RX}}^H(p) \mathbf{H}_{p'}\mathbf{w}_{\mathrm{TX}}(p) {c}_{k,l} + z_{k,l},
\end{align}}
where ${c}_{k,l}$ is the corresponding transmitted symbol, and $\mathbf{H}_{p'}\in\mathbb{C}^{N\times N}$ is the baseband channel response corresponding to the reflection from $p'$, given as
{\small\begin{align}\label{eq:chanHp}
 \mathbf{H}_{p'} &= \mathbf{G}_{u,p',u'}\sqrt{\frac{\sigma_{p'} \lambda^2}{(4\pi)^3 d_{u,p'}^{\alpha}d_{p',u'}^{\alpha}}}e^{j2\pi f_{D,p'}T_o l}e^{-j2\pi \tau_{p'} \Delta f k} 
\end{align} }
with $\mathbf{G}_{u,p',u'}\in\mathbb{C}^{N\times N}$ defined as $\mathbf{G}_{u,p',u'}= \mathbf{g}_{\mathrm{RX}}(\varphi_{p',u'})\mathbf{g}_{\mathrm{TX}}^H(\varphi_{u,p'})$, where $\mathbf{g}_{\mathrm{RX}}(\varphi_{p',u'})\in\mathbb{C}^{N\times 1}$ is the receive beam-steering vector of UAV $u'$ in the direction of cell $p'$ and  $\mathbf{g}_{\mathrm{TX}}(\varphi_{u,p'})\in\mathbb{C}^{N\times 1}$ is the transmit beam-steering vector of UAV $u$ in the direction of cell $p'$. Here, $z_{k,l}$ is the AWGN noise affecting the symbol, $\alpha$ is the pathloss exponent, 
$d_{u,p'}$ is the distance from $u$ to $p'$ and $d_{p',u'}$ is the distance from $p'$ to $u'$. A line-of-sight (LoS) channel is considered, since the performance of sensing is generally dependent on the LoS link between the UAVs and the target, while non-LoS (NLoS) links (if any) are treated as interference for the target sensing \cite{art:Meng_TWC}. In this case, it is considered that UAVs present a strong LoS component, which is favorable for sensing~\cite{art:Meng_TWC}, thus a free-space pathloss model is assumed with $\alpha=2$.

Let $\eta_{p}$ be defined such that $|\eta_{p}| = \sqrt{\sigma_p}$. Assuming zero Doppler from the ground, the combined channel response from the whole area can be generalized to
{\small\begin{align}\label{eq:Htot}
    \mathbf{H} &= \sum_{p'=1}^{P} \mathbf{G}_{u,p',u'}    \frac{\eta_{p'}\lambda e^{-j2\pi \tau_{p'} \Delta f k}}{(4\pi)^{3/2} d_{u,p'}^{\alpha/2}d_{p',u'}^{\alpha/2}}.
\end{align}}

As $\sigma_{\mathrm{G}}$ is uniformly spread over $S$, it follows that $\sqrt{\sigma_{\mathrm{G}}}$ is also uniformly spread across $S$
. Then, for a continuous area, \eqref{eq:Htot} can be extended as { \small
\begin{align}\label{eq:H_int}
    \mathbf{H} &= \frac{\lambda\sqrt{\sigma_{\mathrm{G}}}}{(4\pi)^{3/2} \ell^2}\int\displaylimits_{0}^{\ell} \int\displaylimits_{0}^{\ell}  \mathbf{G}_{u,p',u'}    \frac{e^{-j2\pi \tau_{p'} \Delta f k}}{ d_{u,p'}^{\alpha/2}d_{p',u'}^{\alpha/2}}dx'dy',
\end{align} }
where $dx'$ and $dy'$ are infinitesimals on the $x$ and $y$ coordinates of point $p'$. Similarly, the channel matrix corresponding to the reflections from within the cell defined by $p$ is given as { \small
\begin{align}\label{eq:H_pcell}
    \mathbf{H}_{\{p\}} &= \frac{\lambda\sqrt{\sigma_{\mathrm{G}}}}{(4\pi)^{3/2} \ell^2}\int\displaylimits_{y_{p'}-\frac{d}{2}}^{y_{p'}+\frac{d}{2}} \int\displaylimits_{x_{p'}-\frac{d}{2}}^{x_{p'}+\frac{d}{2}}  \mathbf{G}_{u,p',u'}    \frac{e^{-j2\pi \tau_{p'} \Delta f k}}{ d_{u,p'}^{\alpha/2}d_{p',u'}^{\alpha/2}}dx'dy',
\end{align} }
such that $ \mathbf{H} = \sum_{p=1}^{P} \mathbf{H}_{\{p\}} $. Expressions \eqref{eq:H_int} and \eqref{eq:H_pcell} depend only on the geometry of the system and on the choice of transmit and receive UAV $u$ and $u'$, so it can be computed offline once for every pair of UAVs. By defining $\mathbf{H}_{\{\Bar{p}\}} = \mathbf{H} - \mathbf{H}_{\{p\}}$ and removing the data-dependency from \eqref{eq:symbol_0} by dividing over the transmitted symbol as $\Bar{c}_{k,l} = \Hat{c}_{k,l} / {c}_{k,l}$, the received processed symbols are given as
{\small\begin{align}
 \nonumber   \Bar{c}_{k,l} &=   \underbrace{\mathbf{w}_{\mathrm{RX}}^H(p) \mathbf{H}_{\{p\}}\mathbf{w}_{\mathrm{TX}}(p)}_{\text{Reflections from intended cell}} + \underbrace{\mathbf{w}_{\mathrm{RX}}^H(p) \mathbf{H}_{\{\Bar{p}\}}\mathbf{w}_{\mathrm{TX}}(p)} _{\text{Interferent reflections}}\\
    & + \underbrace{\mathbf{w}_{\mathrm{RX}}^H(p) \mathbf{H}_{p^*}\mathbf{w}_{\mathrm{TX}}(p)}_{\text{Reflection from target}} + \Tilde{z}_{k,l} 
\end{align}}
where $\Tilde{z}_{k,l}$ is the processed noise sample and the reflection from the target has been included, represented by index $p^*$.


\section{Distributed Sensing Protocol}\label{sec:Protocol}

The proposed distributed sensing protocol is summarized in the steps described below.

\textbf{Step 1 (Coordination):} The FC internally constructs the mixed grid, and coordinates the UAVs to assume their positions to cover the whole area of interest $S$. The FC also coordinates the UAVs such that each $u\in\mathcal{U}$ illuminates a set of cells $\mathcal{P}_u$, while the rest obtain and process the corresponding reflections. All cells are illuminated at least once, i.e. $\forall p\in\mathcal{P},~ \exists u\in\mathcal{U},~ s.t. ~ p\in\mathcal{P}_u ~~ \land ~~ \mathcal{P} =  \bigcup\limits_{u\in\mathcal{U}}\mathcal{P}_u$

\begin{figure}[bt]
    \centering
    \includegraphics[width=0.8\linewidth]{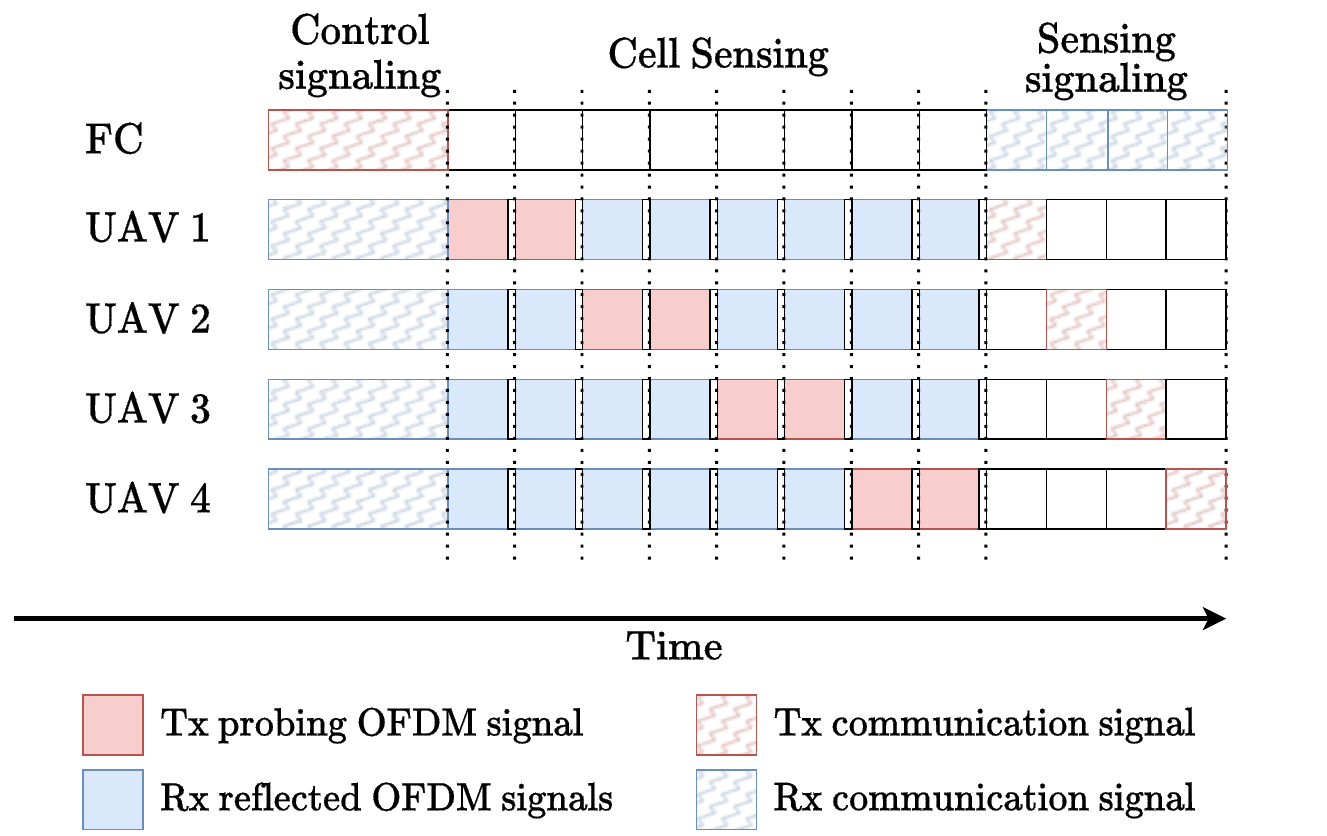}
    \caption{UAV scheduling example.}
    \label{fig:schedule}
\end{figure}

The UAVs are coordinated to follow a schedule in time. This coordination is realized by the FC, indicating the time slots corresponding to the sensing of each cell in the ground, and which UAVs illuminate or receive reflections from it. The schedule also indicates the time slots in which each UAV will send their local statistics to the FC in order to avoid interference between UAVs. An illustration of this scheduling is shown in Fig.~\ref{fig:schedule}.

    \textbf{Step 2 (Digital beamforming):} 
The transmit and receive beamformers to sense cell $p$ illuminated by UAV $u$ and processed by UAV $u'$ are designed to minimize the response of the reflections from the interfering cells while maintaining the main lobe of the transmit beampattern pointing in the direction of the intended cell center. This problem can be expressed as {\small
\begin{subequations}\label{eq:optP_m1}
\begin{alignat}{3}\label{eq:optP_m1_obj}
\mathrm{\textbf{P}:}\;\;\; &\min_{\mathbf{w}_{\mathrm{TX}},\mathbf{w}_{\mathrm{RX}}} & &\mathbf{w}_{\mathrm{RX}}^H(p) \mathbf{H}_{ 
 \{\Bar{p}\}}\mathbf{w}_{\mathrm{TX}}(p) \mathbf{w}_{\mathrm{TX}}(p)^H\mathbf{H}_{\{\Bar{p}\}}^H\mathbf{w}_{\mathrm{RX}}(p) & \\ 
\label{eq:optP_m1_c1} &\text{subject to} &\quad&  \mathbf{w}_{\mathrm{RX}}^H(p)\mathbf{g}_{\mathrm{RX}}(\varphi_{p,u'}) = 1 & \\
\label{eq:optP_m1_c2} & &\quad&  \mathbf{w}_{\mathrm{TX}}^H(p)\mathbf{g}_{\mathrm{TX}}(\varphi_{u,p}) = \sqrt{NP_T} & \\
\label{eq:optP_m1_c3} & &\quad&  || \mathbf{w}_{\mathrm{TX}} ||_2^2 \leq P_T. &
\end{alignat}
\end{subequations}}
where constraint \eqref{eq:optP_m1_c1} provides distortionless response in reception, constraint \eqref{eq:optP_m1_c2} gives the maximum response in the desired direction by employing the Cauchy-Schwarz inequality, and constraint \eqref{eq:optP_m1_c3} provides an upper-bound of the transmit power $P_T$. To fulfill \eqref{eq:optP_m1_c2} while enforcing \eqref{eq:optP_m1_c3}, the transmit beamformer should be expressed as
{\small\begin{align}\label{eq:txBF}
    \mathbf{w}_{\mathrm{TX}}(p) = \sqrt{\frac{P_T}{N}}\mathbf{g}_{\mathrm{TX}}(\varphi_{u,p}).
\end{align}}

Then, by fixing the transmit beamformer, problem \textbf{P} gives the minimum variance distortionless response receive beamformer given as~\cite{art:Stoica_Capon} 
{\small\begin{align}\label{eq:rxBF}
    \mathbf{w}_{\mathrm{RX}}(p) = \frac{\mathbf{R}_{p,u'}^{-1}\mathbf{g}_{\mathrm{RX}}(\varphi_{p,u'})}{\mathbf{g}_{\mathrm{RX}}(\varphi_{p,u'})^H\mathbf{R}_{p,u'}^{-1}\mathbf{g}_{\mathrm{RX}}(\varphi_{p,u'})}.
\end{align}}
where $\mathbf{R}_{p,u'} = \mathbf{H}_{\{\Bar{p}\}}\mathbf{w}_{\mathrm{TX}}(p) \mathbf{w}_{\mathrm{TX}}(p)^H\mathbf{H}_{\{\Bar{p}\}}^H + \kappa \mathbf{I}_N$, with $\mathbf{R}_{p,u'}\in\mathbb{C}^{N\times N}$, $\mathbf{I}_N\in\mathbb{R}^{N\times N}$ as the identity matrix and $\kappa$ is a small real number~\cite{art:Shi_ILS}.

As defined above, the transmit beamforming done at the transmitting UAV depends only on the center point of the cell to illuminate. However, the receive beamforming is considered to depend on the matrix $\mathbf{H}_{\{\Bar{p}\}}$, which accounts for the total continuous reflections from the ground outside the intended cell. The process of computing such a matrix may be strenuous at the receive UAV depending on the meshing used for integration. Therefore, it is proposed an alternative simplified receive reflection matrix, which is expressed as
{\small\begin{align}\label{simplemat}
    \Hat{\mathbf{H}}_{\{\Bar{p}\}} = \frac{\lambda\sqrt{\sigma_{\mathrm{G}}}}{(4\pi)^{3/2} \ell^2} \sum_{\substack{p'=1\\p'\neq p}}^{P}\mathbf{G}_{u,p',u'}    \frac{e^{-j2\pi \tau_{p'} \Delta f k}}{ d_{u,p'}^{\alpha/2}d_{p',u'}^{\alpha/2}}d^2,
\end{align}}
In \eqref{simplemat}, the interference is assumed as reflected only from the center of every cell, except from the intended sensed cell $p$.

\textbf{Step 3 (Distributed sensing):} 
After the UAVs compute the transmit and receive beamformers corresponding to all cells $p\in\mathcal{P}$, the schedule coordinated by the FC is enforced. At the corresponding time, UAV $u\in\mathcal{U}$ sequentially illuminates its assigned cells $p\in\mathcal{P}_u$ by applying the associated transmit beamformers $\mathbf{w}_{\mathrm{TX}}(p)$. Every other UAV  $u'\in\mathcal{U}\setminus\{u\}$ gathers the incoming reflections by employing the corresponding receive beamformer $\mathbf{w}_{\mathrm{RX}}(p)$, sampling the reflections at the expected delay for the cell $\tau_p$, and obtaining the received symbols from the OFDM frames over all the subcarriers. Once the schedule is complete, every cell has been illuminated, and every UAV has obtained the symbols $\Hat{c}_{k,l}$ for all $l\in\{0,1,...,M_s-1\}$ and $k\in\{0,1,...,N_c-1\}$ corresponding to the cells $p\in\mathcal{P}\setminus\mathcal{P}_u$. This occurs because the UAVs work in half-duplex mode, not being able to obtain the reflections coming from their own illuminated cells.

\textbf{Step 4 (Local processing): } Each UAV prepares certain statistics of their locally received signals, which depend on the estimation method used, to be sent to the FC for central processing. There are three methods that are considered for obtaining such statistics, namely MIMO RCS MLE estimation (MIMORE), multistatic RCS MLE estimation (MuRE) and multistatic position estimation (MuPE). 
First, statistics employed for MLE are developed from the received signal at the UAVs, and then the three mentioned methods that make use of such statistics are introduced.
Henceforth, the superscript $\cdot^{(u')}$ represents the corresponding quantities from the perspective of receive UAV $u'$.

Let UAV $u'$ receive the reflections coming from intended sensed cell $p$. Given that the interference from the ground is known, it can be removed before processing, while any unknown interference coming from OFDM signals can be modeled as an additional source of gaussian noise~\cite{Chen_JSACUAVSystem}. Then, the received symbol corresponding to the $l$-th OFDM symbol and the $k$-th subcarrier is given as
{\small\begin{align}
\nonumber    \!\!\Hat{c}^{(u')}_{k,l} \!\!\!\! &=   \mathbf{w}_{\mathrm{RX}}^H(p) \mathbf{H}_{p}\mathbf{w}_{\mathrm{TX}}(p)c^{(u)}_{k,l} + \Hat{z}_{k,l}\\
   \!\! &= \!\! \frac{\chi_{\mathrm{RX}}\chi_{\mathrm{TX}}^* \eta_p\lambda}{(4\pi)^{\frac{3}{2}} d_{u,p'}^{\alpha/2}d_{p',u'}^{\alpha/2}} e^{j2\pi f_{D,p'}T_o l}e^{-j2\pi \tau_{p'} \Delta f k}c^{(u)}_{k,l}\!\!\!+\! \Hat{z}_{k,l}
\end{align}}
where $\chi_{\mathrm{RX}} $$=$$ \mathbf{w}_{\mathrm{RX}}^H(p)\mathbf{g}_{\mathrm{RX}}(\varphi_{p,u'})$, $\chi_{\mathrm{TX}} = \mathbf{w}_{\mathrm{TX}}^H(p)\mathbf{g}_{\mathrm{TX}}(\varphi_{p,u})$ and $\Hat{z}_{k,l}$ is the AWGN noise component that includes unknown interference. Let $\mathbf{\Hat{c}}^{(u')}\in\mathbb{C}^{M_sN_c\times 1}$ and $\mathbf{c}^{(u)}\in\mathbb{C}^{M_sN_c\times 1}$ be the arrays of vectorized receive and transmit symbols. The phase shift vectors $\mathbf{v}^{(u')}_{f}\in\mathbb{C}^{N_c\times 1}$ and $\mathbf{v}^{(u')}_{\tau}\in\mathbb{C}^{M_s\times 1}$ are defined as
$ \mathbf{v}^{(u')}_{f} = [1,e^{-j2\pi f^{(u')}_{D,p}T_o(1)},...,e^{-j2\pi f^{(u')}_{D,p}T_o(N_c-1)}]^T $ and $ \mathbf{v}^{(u')}_{\tau} = [1,e^{j2\pi \tau^{(u')}_{p} \Delta f(1)},...,e^{j2\pi \tau^{(u')}_{p} \Delta f(M_s-1)}]^T $. 
The total phase shift vector corresponding to UAV $u'$ is given as $\mathbf{v}^{(u')}=\mathbf{v}^{(u')}_{\tau}\otimes\mathbf{v}^{(u')}_{f}$ with $\mathbf{v}^{(u')}\in\mathbb{C}^{M_sN_c\times 1}$. The vectors of received and transmitted symbols corresponding to the cell $p$ obtained by every UAV $u'\in\mathcal{P}\setminus\mathcal{P}_u$ are given by $\mathbf{\Hat{c}} = [(\mathbf{\Hat{c}}^{(1)})^T,...,(\mathbf{\Hat{c}}^{(U)})^T]^T$ and $\mathbf{c} = \mathbf{c}^{(u)}\otimes \mathbf{1}_{U-1}$, with $\mathbf{c},\mathbf{\Hat{c}}\in\mathbb{C}^{(U-1)M_sN_c\times 1}$. Similarly, the total vector of phase shifts normalized with respect to the corresponding pathloss is $\mathbf{v} = [(d_{p,1}^{-\alpha/2}\mathbf{v}^{(1)})^T,...,(d_{p,U}^{-\alpha/2}\mathbf{v}^{(U)})^T]^T$, with $\mathbf{v}\in\mathbb{C}^{(U-1)M_sN_c\times 1}$. Considering constraints \eqref{eq:optP_m1_c1} and \eqref{eq:optP_m1_c2}, the central received symbol vector is given as
{\small\begin{align}
    \mathbf{\Hat{c}} &= \underbrace{\frac{\sqrt{NP_T} \lambda}{(4\pi)^{\frac{3}{2}} d_{u,p}^{\alpha/2}}\eta_p(\mathbf{v} \odot \mathbf{c})}_{\mathbf{s}} + \mathbf{z}.
\end{align}}
Here, $\mathbf{z}\in\mathbb{C}^{(U-1)M_sN_c\times 1}$ is the vector of noise samples, and is a circularly symmetric gaussian vector of mean zero and covariance matrix $\mathbf{C}=N_0 \mathbf{I}_{N_cM_s(U-1)}$. Then, the likelihood function is given as
{\small\begin{align}
\mathcal{L}(\mathbf{\Hat{c}};\eta_p) &= \frac{1}{(\pi N_0)^{M_sN_c(U-1)}}e^{-\frac{1}{N_0}||\mathbf{\Hat{c}}-\mathbf{s}||_2^2}.
\end{align}}
Discarding terms that do not depend on $\eta_p$, the log-likelihood function is obtained for ease of optimization, as a monotonically increasing function that preserves the concavity of the likelihood function, and is given as
{\small\begin{align}
    l(\mathbf{\Hat{c}};\eta_p) &= - \frac{1}{N_0}\left(||\mathbf{s}||_2^2-2\mathcal{R}\{ \mathbf{s}^H \mathbf{\Hat{c}} \}\right)
\end{align}}

Taking the Wirtinger derivatives over $\eta_p$ and equaling the expression to zero, the MLE for the RCS is obtained as in \eqref{eq:MLE_RCS}, at the bottom of next page. If data dependency is removed from the received symbols prior to processing, as $\Bar{c}^{(u')}_{k,l} = \Hat{c}^{(u')}_{k,l} / c^{(u')}_{k,l}$, the MLE for the RCS is simplified to \eqref{eq:MLE_RCS_2}, at the bottom of the next page. These expressions are used to obtain the statistics to be computed by each receive UAV and later sent to the FC for fusion, depending of the estimation method used. These methods are introduced next.

\begin{figure*}[b]{\small
\begin{align}\label{eq:MLE_RCS}
    \Hat{\sigma}_p &= \frac{(4\pi)^3 d_{u,p}^{\alpha}}{NP_T\lambda^2} \left(\sum\limits_{\substack{u'=1 \\ u'\neq u}}^{U} d_{p,u'}^{-\alpha} \sum\limits_{l=0}^{N_c-1}\sum\limits_{k=0}^{M_s-1}| c^{(u)}_{k,l} |^2\right)^{-2} \left| \sum\limits_{\substack{u'=1 \\ u'\neq u}}^{U} d_{p,u'}^{-\alpha/2} \sum\limits_{l=0}^{N_c-1}\sum\limits_{k=0}^{M_s-1} (c^{(u)}_{k,l})^*\Hat{c}^{(u')}_{k,l} e^{-j2\pi f_{D,p}^{(u')}T_o l}e^{j2\pi \tau_{p}^{(u')} \Delta f k}  \right|^2 \\ 
\label{eq:MLE_RCS_2}    \Hat{\sigma}_p &= \left(\frac{1}{M_sN_c}\right)^2\frac{(4\pi)^3 d_{u,p}^{\alpha}}{NP_T\lambda^2} \left(\sum\limits_{\substack{u'=1 \\ u'\neq u}}^{U} d_{p,u'}^{-\alpha} \right)^{-2}\left| \sum\limits_{\substack{u'=1 \\ u'\neq u}}^{U} d_{p,u'}^{-\alpha/2} \sum\limits_{l=0}^{N_c-1}\sum\limits_{k=0}^{M_s-1} \Bar{c}^{(u')}_{k,l} e^{-j2\pi f_{D,p}^{(u')}T_o l}e^{j2\pi \tau_{p}^{(u')} \Delta f k}  \right|^2
\end{align}}
\end{figure*}


\textbf{\textit{MIMORE:} }
Under this approach, the RCS of every cell $p\in\mathcal{P}$ is centrally computed as in \eqref{eq:MLE_RCS_2}. For this, each UAV $u'\in\mathcal{U}$ sends their statistics $\delta^{(u')}_p$ defined as
{\small\begin{align}\label{eq:MIMORE_stat}
    \delta^{(u')}_p = \sum\limits_{l=0}^{N_c-1}\sum\limits_{k=0}^{M_s-1} \Bar{c}^{(u')}_{k,l} e^{-j2\pi f_{D,p}^{(u')}T_o l}e^{j2\pi \tau_{p}^{(u')} \Delta f k},
\end{align}}
for every cell $p\in\mathcal{P}\setminus\mathcal{P}_u$ to the FC for central estimation. From \eqref{eq:MLE_RCS_2}, these are sufficient statistics for the central RCS estimation of $p$.

\textbf{\textit{MuRE:} }
Under this approach, the RCS of every cell $p\in\mathcal{P}\setminus\mathcal{P}_u$ is locally computed by each UAV $u'\in\mathcal{U}$, forming a local RCS map of the grid $\mathbf{\Hat{\Gamma}}_{u}$. The UAVs then send their local RCS maps to the FC for central estimation. Considering the MLE RCS estimation of a single receive UAV $u'$, equation \eqref{eq:MLE_RCS_2} becomes
{\small\begin{align}
    \nonumber \Hat{\sigma}_p &= \left(\frac{1}{M_sN_c}\right)^2\frac{(4\pi)^3 d_{u,p}^{\alpha}d_{p,u'}^{\alpha}}{NP_T\lambda^2}\times \\ \label{eq:MLE_Local}
    &\left| \sum\limits_{l=0}^{N_c-1}\sum\limits_{k=0}^{M_s-1} \Bar{c}^{(u')}_{k,l} e^{-j2\pi f_{D,p}^{(u')}T_o l}e^{j2\pi \tau_{p}^{(u')} \Delta f k}  \right|^2.
\end{align}}
Then, each receive UAV estimates the RCS of each of the cells $p\in\mathcal{P}\setminus\mathcal{P}_u$ locally with \eqref{eq:MLE_Local}, obtaining 
$\mathbf{\Hat{\Gamma}}_{u}$. 
Note that $\mathbf{\Hat{\Gamma}}_u$ is a matrix of RCS estimates of all cells $p\in\mathcal{P}$, where the estimates for the cells $p\in\mathcal{P}_u$ are set to zero. 


\textbf{\textit{MuPE:} } Each UAV $u\in\mathcal{U}$ obtains local estimates of the RCS of every cell $p\in\mathcal{P}$ employing \eqref{eq:MLE_Local}, obtaining an RCS map of the grid $\mathbf{\Hat{\Gamma}}_{u}$. Afterwards, each UAV locally estimates the position of the target $\Hat{\mathbf{r}}_{p^*}^{(u)}$ as described in Section~\ref{sec:posEst}. Under this approach, the $\Hat{\mathbf{r}}_{p^*}^{(u)}$ estimates are sent to the FC for fusion.

\textbf{Step 5 (Centralized estimation):} In this stage the statistics gathered from every UAV $u'\in\mathcal{U}$ are processed at the FC to obtain the central estimate of the location of the target. According to the different approaches introduced in the previous step, there are three different methods for processing at the FC, depending on the statistics gathered at the UAVs. These are described below.

\textbf{\textit{MIMORE: }} The FC gathers the statistics $\delta^{(u')}_p$ for each UAV $u'\in\mathcal{U}$ and every cell $p\in\mathcal{P}\setminus\mathcal{P}_u$, Then, the RCS of every cell $p\in\mathcal{P}$ is estimated as in \eqref{eq:MLE_RCS_2}.
With these estimates, the FC obtains the central RCS map of the grid $\mathbf{\Hat{\Gamma}}$. Finally, the FC estimates the position of the target $\Hat{\mathbf{r}}_{p^*}^{(u)}$ as described in Section~\ref{sec:posEst}.

\textbf{\textit{MuRE: }} The FC gathers the local RCS maps $\mathbf{\Hat{\Gamma}}_u$ of all UAVs $u\in\mathcal{U}$ and performs information-level fusion of the local estimates to obtain a global estimate $\mathbf{\Hat{\Gamma}}$. To this end, the FC averages the values of each cell over the local maps from all UAVs in $\mathcal{U}$ such that $\mathbf{\Hat{\Gamma}} = \frac{1}{U}\sum_{u\in\mathcal{U}} \mathbf{\Hat{\Gamma}}_{u}$. 
Then, the FC estimates the position of the target based on the fused RCS map, as described in Section~\ref{sec:posEst}.

\textbf{\textit{MuPE: }} The FC gathers the local position estimates $\Hat{\mathbf{r}}_{p^*}^{(u)}$ of all UAVs $u\in\mathcal{U}$ and performs information-level fusion of the local estimates of the position of the target by averaging them as $\Hat{\mathbf{r}}_{p^*} = \frac{1}{U}\sum_{u\in\mathcal{U}} \Hat{\mathbf{r}}_{p^*}^{(u)}$.

\subsection{Overhead Analysis}

Assume that $\Hat{\sigma_p}\in\mathbf{\Hat{\Gamma}}_{u}$ is represented by a 32-bit floating point number, $\Hat{\mathbf{r}}_{p^*}^{(u)}$ is represented by two 32-bit floating point numbers (one for each cartesian component), while $\delta^{(u')}_p$ is represented by two 32-bit floating point numbers (real and imaginary components). The overhead sent by each UAV to the FC when reporting their local RCS maps $\mathbf{\Hat{\Gamma}}_{u}$, is of $32P$ bits. When reporting their local statistics $\delta^{(u')}_p$, the overhead sent by each UAV to the FC is of $64P$ bits. Finally, only $64$ bits are sent by each UAV when reporting their local position estimates $\Hat{\mathbf{r}}_{p^*}^{(u)}$. 

For comparison, a CS technique~\cite{art:Heath_PMNFramework} for the central estimation of the target location would require a total of $M_sN_cN$ samples per cell to be sent to the FC by each UAV. Each of those samples would be represented by two 32-bit floating point numbers (real and imaginary components), so that the payload sent by each UAV to the FC when reporting their samples for CS would be of $64M_sN_cNP$ bits. As an illustrative example, assuming a base grid of $12\times 12$ cells and $16$ UAVs deployed uniformly across the grid as in Fig.~\ref{fig:mixedGrid}, with OFDM frames consisting of 64 subcarriers and 16 OFDM symbols, and UPAs of 4 antenna elements, the overhead for the transmission of a single UAV, and the total reception overhead of the FC are shown in Table~\ref{tab:load_analysis}, by considering a mixed grid. 
It is worthwhile to notice that there is a remarkable overhead reduction attained by the proposed algorithms if compared to the CS approach.

\begin{table}[bt]\vspace{-1em}\centering
\caption{Overhead analysis}\label{tab:load_analysis}
  \begin{tabular}{|c|c|c|c|c|}
    \hline
                        &  \textbf{CS}  & \textbf{MIMORE}& \textbf{MuRE}       &   \textbf{MuPE}  \\ \hline
    Tx    &   50069 [Kbits]          & 12.22 [Kbits]  & 6.11 [Kbits]        &   0.064 [Kbits]         \\ \hline
    Rx     &   801112 [Kbits]          & 195.58 [Kbits] & 97.79 [Kbits]       &   1.024 [Kbits]           \\ \hline
  \end{tabular}
   \vspace{-1em}
\end{table}


\section{Position Estimation}\label{sec:posEst}

The estimation of the position of the target is done locally by each UAV or centrally by the FC, depending on the considered method.
To that purpose, an on-grid approach is employed, where the target position estimate $p^*$ is restricted to be the center point of the cell with highest estimated RCS on the grid, disregarding the information of adjacent cells. To improve the accuracy of this method, an off-grid refinement technique is also employed. 

\subsection{On-Grid Estimation}

The on-grid estimation considers that 
the target is located at the cell 
that returns the maximum RCS estimate $\Hat{\sigma}_{p'}=\max_{p\in\mathcal{P}} \mathbf{\Hat{\Gamma}}$.
Therefore, the on-grid estimation of the position of the target is $\Hat{\mathbf{r}}_{p^*} = \mathbf{r}_{p'}$. Note that if the cell containing the target is correctly estimated, the $L^\infty$ distance estimation error is bounded by half of the size of the cell $d$ as $0\leq || \Hat{\mathbf{r}}_{p^*}- \mathbf{r}_{p^*} ||_\infty \leq d/2$.
To reduce the off-grid estimation error, off-grid post-processing of the estimate is carried out as explained below.

\subsection{Off-Grid Post-Processing}

The off-grid post-processing over the on-grid estimates is performed by considering a weighted average technique as in~\cite{art:Feng_postproc}. For this purpose, all elements of $\mathbf{\Hat{\Gamma}}$ are normalized between 0 and 1, thus, the set of all cells in $\mathbf{\Hat{\Gamma}}$ that have a value above a certain threshold $\lambda$ is defined as $\mathcal{T} = \{ p | p\in \mathbf{\Hat{\Gamma}} ; \Hat{\sigma_p} \geq \lambda \}$. Then, the off-grid estimate $\Hat{\mathbf{r}}_{p^*}$ is given by the weighted average of the positions of all elements in $\mathcal{T}$ as $\Hat{\mathbf{r}}_{p^*} = \sum_{p\in \mathcal{T}} \alpha_p \mathbf{r}_{p}$,
with $\alpha_p = \frac{\Hat{\sigma_p}}{\sum_{p'\in \mathcal{T}} \Hat{\sigma_{p'}}}$.


\section{Numerical Results}\label{sec:Results}

The performance of the proposed target localization framework is evaluated in terms of the absolute error between the estimated and the real target position, given as $ || \Hat{\mathbf{r}}_{p^*} - \mathbf{r}_{p^*} ||_2$, which is the Euclidean norm of the estimation error. For this purpose, Monte Carlo simulations are performed, where the target is randomly and uniformly located in the area of interest at each iteration. The performance of the proposed methods is evaluated for the mixed grid approach. These results are compared to a full base-grid approach. Then, when the value of $\lambda = 1$ is considered, just the cell of highest value is selected, while for a value of $\lambda=0.9$ the highest valued cells are considered for estimation. In addition, the results are compared to the case where a single UAV is performing sensing by operating in full-duplex mode from the geometric center of the area of interest. The simulation parameters utilized for all figures are presented in Table~\ref{tab:commonPar}, unless stated otherwise. 
\begin{table}[bt]\vspace{-1em}\centering
\caption{Common simulation parameters}\label{tab:commonPar}
  \begin{tabular}{|c|c||c|c|}
    \hline
    \textbf{Parameter}  & \textbf{Value}    &   \textbf{Parameter}  & \textbf{Value}       \\ \hline
    $M_s$                   & 16             &      $N_0$                   & -109 [dBm]      \\ \hline
    $N_c$                   & 64             &      $h$                   & 100 [m]         \\ \hline
    $f_0$                   & 24 [GHz]       &      $L$                     & 18     \\ \hline
    $BW$                    & 200 [MHz]      &      $U$                     & 9     \\ \hline
    $P_T$                   & 30 [dBm]         &      $\sigma_{\mathrm{G}}$   & 25 [$m^2$]     \\ \hline
    $N$                     & 16             &      $\sigma_{\mathrm{T}}$   & 10 [$m^2$]     \\ \hline
    $\ell$                     & 50 [m]             &      $f_D$   & 0 [Hz]     \\ \hline
  \end{tabular}
   \vspace{-1em}
\end{table}

\begin{figure}[ht]
    \centering
    \includegraphics[width=0.8\linewidth]{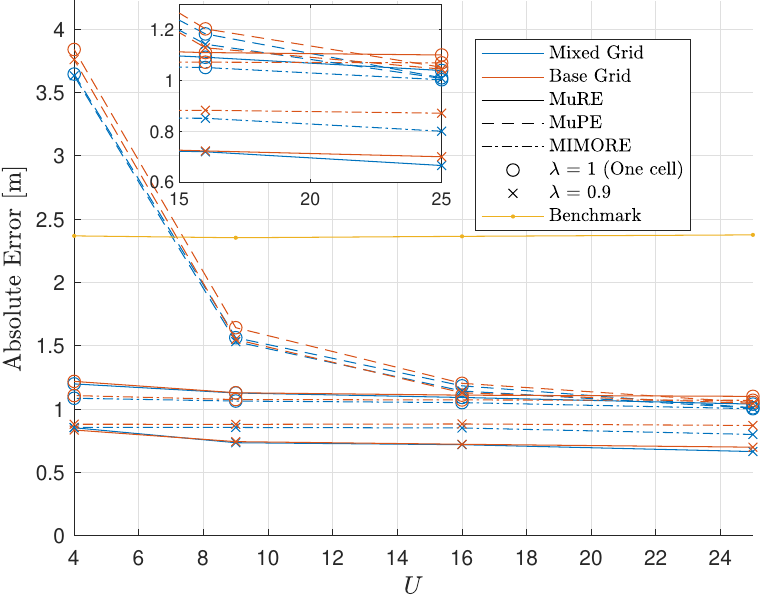}
    \caption{Absolute error of the estimation of the position of the target for different number of UAVs $U$ for different types of estimation.\vspace{-1em}}
    \label{fig:resultN_nUAV}
\end{figure}

Fig.\ref{fig:resultN_nUAV} presents the absolute estimation error plotted versus number of UAVs in the system, $U$. Note that, for all the proposed techniques, increasing the number of UAVs decreases the estimation error for the position of the target. 
It is worth observing that the employment of multiple UAVs in the proposed system provides several advantages, such as having more samples to estimate the RCS of a cell, any delay ambiguity due to the MLE estimation is reduced, and the pathloss is reduced as closer UAVs are available. It can be observed that the MuPE case present higher errors for smaller number of UAVs even when compared to the benchmark. This is explained by the fact that, in the MuPE case, each UAV estimates its own target position. The UAV that directly illuminates the target does not hear reflections due to the half-duplex operation, thus its target position estimate is more biased. Therefore, smaller number of UAVs in the system are more impacted by this biased estimate in average, thus increasing the error in the overall position estimate of the target.  

It is also observed that, when no post-processing is considered ($\lambda=1$), the MIMORE method has better performance than the other methods, which is to be expected, as it employs the maximum likelihood estimator in a distributed setting, ensuring better results than estimates based on local estimation. With post-processing, i.e. $\lambda=0.9$, the performance of every case improves by further lowering the estimation error, and the MuRE case presents the best performance. Moreover, the mixed grid approach presents better performance than the base grid approach in every case, because of the increased sampling around the edges between $\mathcal{P}_u$ sets of cells.


\begin{figure}[ht]
    \centering
    \includegraphics[width=0.8\linewidth]{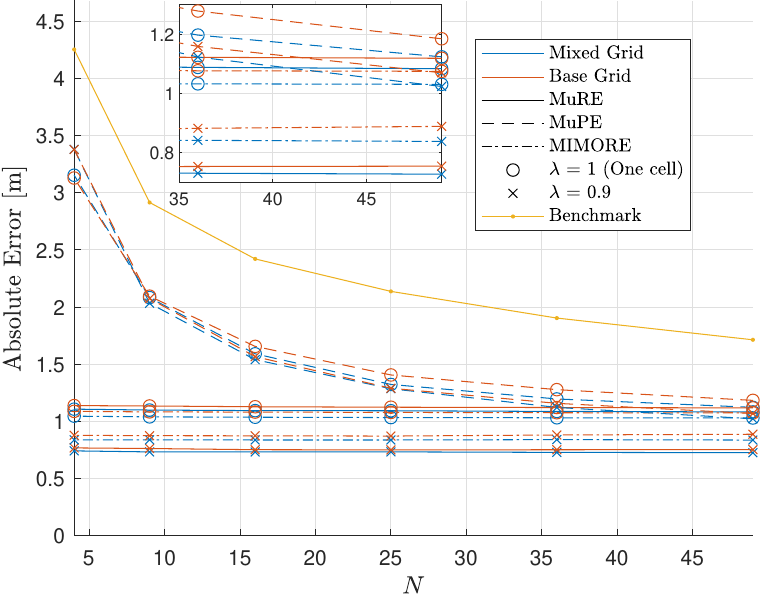}
    \caption{Absolute error of the estimation of the position of the target for varying number of antennas $N$ in the UPAs of the UAVs. \vspace{-1em}}
    \label{fig:resultN_nAnt}
\end{figure}

In Fig.~\ref{fig:resultN_nAnt}, the absolute estimation error is plotted versus the number of antennas $N$ in the UPAs of the UAVs. Here, the MIMORE and MuRE methods maintain an almost constant error, while the MuPE method presents a decreasing in estimation error as more antennas are used. 
This effect occurs because, for the MuPE algorithm, there is an ambiguity introduced by MLE estimation, coming from the points in the ground with the same delay as the target, which is reduced only by the beamforming. For this reason, narrower beams allow for smaller ambiguity regions in the ground in the estimation of the RCS of the cells, thus reducing the estimation error. 
In MIMORE and MuRE, this ambiguity is reduced by the multiplicity of UAVs having areas of ambiguity that overlap only at the areas around where the target is located. Then, the non-overlapping areas are decreased when the central estimation is performed. Thus, leading to a more accurate position estimation of the target. The behavior of the other curves is the same as previously seen. 

\begin{figure*}[ht]
    \centering
    \includegraphics[width=0.94\linewidth]{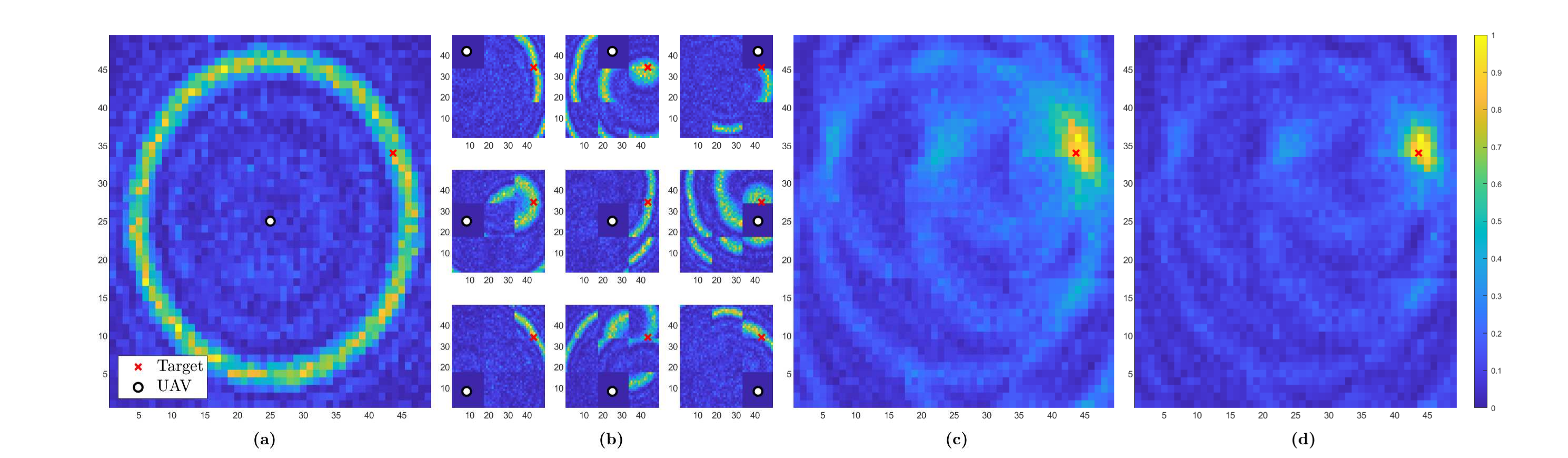}
    \caption{RCS maps with $N=1$ single isotropic antenna element in the UAVs for (a) Single monostatic UAV benchmark, (b) Local RCS maps for distributed UAVs, (c) MuRE central RCS map, (d) MIMORE central RCS map.\vspace{-1em}}
    \label{fig:ambiguity_N1}
\end{figure*}
\begin{figure*}[ht]
    \centering
    \includegraphics[width=0.94\linewidth]{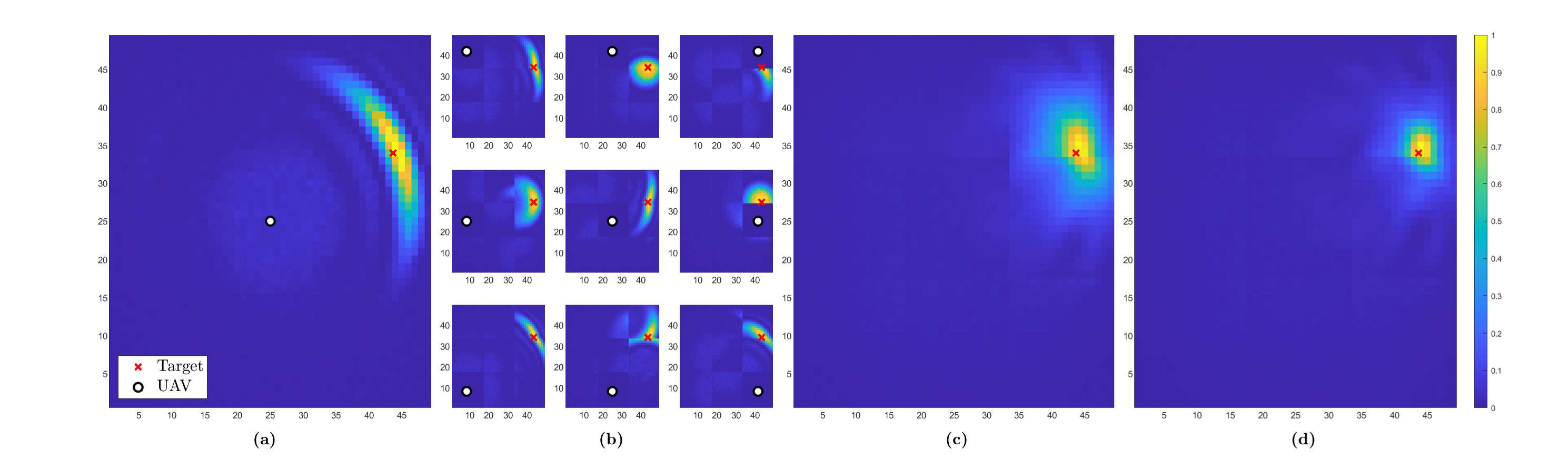}
    \caption{RCS maps with $N=64$ antenna elements in the UAV UPAs for (a) Single monostatic UAV benchmark, (b) Local RCS maps for distributed UAVs, (c) MuRE central RCS map, (d) MIMORE central RCS map.\vspace{-1em}}
    \label{fig:ambiguity_N64}
\end{figure*}

To visualize the ambiguity due to the delay and the reduction of such ambiguity, in Fig.~\ref{fig:ambiguity_N1} and Fig.~\ref{fig:ambiguity_N64}, the RCS maps are shown for the single monostatic UAV benchmark, for all of the distributed UAVs, for the central MuRE algorithm, and for the central MIMORE algorithm. Fig.~\ref{fig:ambiguity_N1} shows the maps for $N=1$ single isotropic antenna for the UAVs and Fig.~\ref{fig:ambiguity_N64} shows the maps for $N=64$. Note that the monostatic and the local distributed RCS maps present an ambiguity in the angular domain around the target, as those are points with the same delay with respect to the transmit and receive UAVs. 

This ambiguity is reduced as the number of antennas increase, as the main lobe of the beams get narrower. However, it is observed that a more significant reduction on the ambiguity occurs when performing centralized estimation in the FC. For algorithm MuRE the ambiguity is highly reduced, and it is even further reduced with the MIMORE algorithm. This behavior is more accentuated when considering a single antenna element in the UAVs, i.e. no beamforming was performed. This effect occurs because, the reflections obtained by each individual UAV have different zones of ambiguity, but all of them share the point where the target is located. So, by performing fusion, the cell corresponding to the position of the target is emphasized, while the ambiguity zones are decreased. 
Looking at the local UAV maps in (b), it can be seen that for $N=1$, the regions of ambiguity are non-negligible and are more likely to impact the performance of the MuRE algorithm (which was also observed in Fig.~\ref{fig:resultN_nAnt}).

\begin{figure}[ht]
    \centering
    \includegraphics[width=0.8\linewidth]{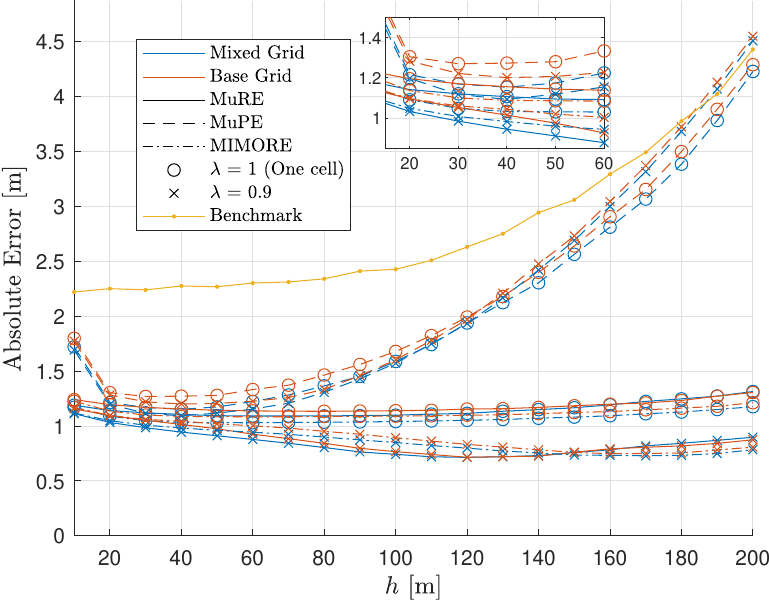}
    \caption{Absolute error of the estimation of the position of the target for varying altitude of the UAVs $h$ for different types of estimation.\vspace{-2em}}
    \label{fig:resultN_hJ}
\end{figure}

In Fig.\ref{fig:resultN_hJ} the absolute estimation error is plotted versus the altitude of the UAVs $h$. It is observed a point of minimum error, for which smaller altitudes negatively impact the performance of the estimators. This behavior occurs due to the beams pointing to the cells are closer to $90^\circ$ in the elevation domain at smaller altitudes, then the beams are ill-behaved at this points. In particular, the MuPE 
method is severely impacted by this effect, as well as the delay ambiguity discussed previously, as the altitude of the UAVs further increase approaching the benchmark. On the other hand, the errors for MIMORE and MuRE remain low even at high altitudes. As in the previous figures, MIMORE presents better performance without post-processing, and post-processing with $\lambda=0.9$ provides a notable reduction of the estimation error, with the best results achieved for the MuRE until 140m, while the best results are achieved by MIMORE above that value.

\begin{figure}[ht]
    \centering
    \includegraphics[width=0.8\linewidth]{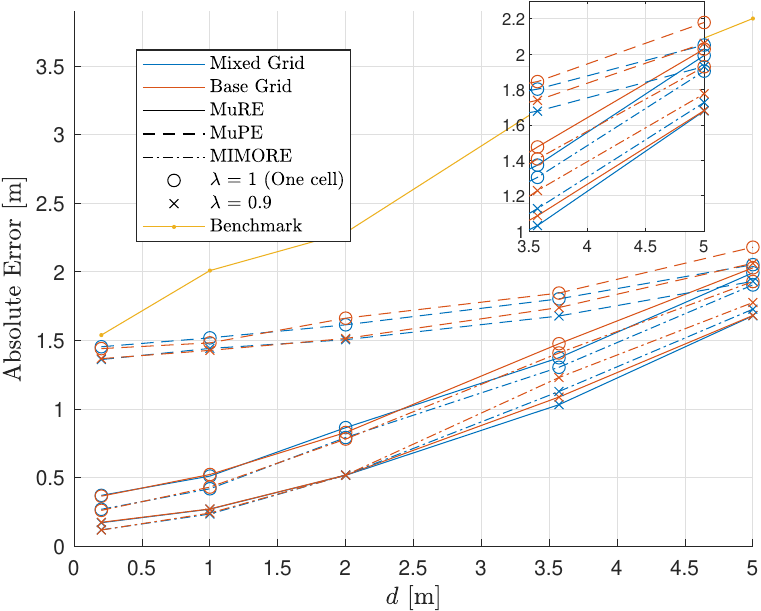}
    \caption{Absolute error of the estimation of the position of the target for varying size of the cells $d$ in meters. \vspace{-2em}}
    \label{fig:resultN_Lam_d}
\end{figure}

In Fig.\ref{fig:resultN_Lam_d} the absolute estimation error is plotted versus the size of the cells $d$ in meters. It is observed that the absolute estimation error monotonically increases with the size of the cells. This behavior is expected as the average minimum error is bounded by the dimensions of the cells. 
For small cells of 20cm, the absolute error is smallest for MIMORE method, presenting an error of 26cm without post-processing and 12cm with $\lambda=0.9$, which is an improvement of less than half of the error with post-processing. The behavior of the other curves is the same as previously seen.

\section{Conclusions}\label{sec:Conclusions}

A half-duplex distributed sensing framework for UAV-assisted networks was proposed, based on the RCS estimation of a spatial on-grid approach. A mixed grid was proposed for on-grid refinement while a partial weighted average post-processing was proposed for off-grid refinement. For RCS estimation three algorithms were proposed and contrasted to benchmarks in terms of the introduced transmission overhead and the absolute error for the estimation of the position of a ground point-like target.



Our results showed that the proposed distributed framework using the MLE algorithms presents advantages with respect to the CS estimation techniques in terms of transmission overhead. This framework also presents a significant improvement with respect to the estimation performance attained by a single monostatic UAV benchmark. For the MuPE method, making the beams narrower increases the performance of the estimation, but for the MIMORE and MuRE methods, this effect is negligible as the ambiguity introduced by the ML estimation is mitigated by the multiplicity of UAVs. For MuRE and MIMORE, having as few as 4 UAVs to cover the area of interest working in half-duplex mode greatly outperforms the benchmark of a single UAV working in full duplex monostatic mode. The MuRE and MIMORE methods are robust against higher UAV altitudes, in which the benchmark presents much higher errors. For small cells of 20cm and considering post-processing, the MIMORE method can reach estimation errors as low as 12cm. In general, the off-grid post-processing reduces the estimation error in all cases, even with a simple choice of $\lambda=0.9$. 
The mixed grid results have lower error in every case due to the increased sampling by a subset of the UAVs. 
\newpage
\bibliographystyle{IEEEbib}\label{sec:refs}
\bibliography{refs}


 




\vfill

\end{document}